\documentclass[journal=jacsat,manuscript=article]{achemso}

\usepackage[version=3]{mhchem} 
\usepackage[T1]{fontenc}       
\usepackage{graphicx}
\usepackage{bm}
\usepackage{multirow}
\usepackage{amssymb}
\usepackage{color}

\usepackage[font=small]{caption}
\usepackage[labelformat=empty, position=top]{subcaption}
\usepackage{graphicx}
\usepackage[export]{adjustbox}

\author{Oscar Y. Fajardo}
\author{Silvia Di Lecce}
\author{Fernando Bresme}
\email{fbresme@imperial.ac.uk}
\affiliation[Imperial College]
{Department of Chemistry, Imperial College London, London UK}
\title[A Generic Coarse-Grained model]
  {Molecular Dynamics Simulation of Imidazolium C$_\mathrm{n}$MIM-BF$_4$ Ionic Liquids using a Coarse Grained Force-Field} 
 
\abbreviations{IL,CG,CnMIM}

\begin{document}

\begin{abstract}
Ionic Liquids feature thermophysical properties that are of interest in solvents,  energy storage materials and tenable lubrication applications. Recently, a series of coarse grained (CG) models was developed to investigate 1-ethyl-3-methylimidazolium tetrafluoroborate [C$_2$MIM][BF$_4$] and 1-butyl-3-methylimidazolium hexafluorophosphate [C$_4$MIM][PF$_6$] ionic liquids. Building on these CG models we derive force-fields to investigate the [C$_{2-8}$MIM] [BF$_4$] family of ionic liquids, as a starting step for systematic investigations of lubrication under nanoconfinement conditions. The simualted equation of state and diffusion coefficients are in good agreement with experimental data and with all-atom force fields. We use these model to analyze the nano 
nanostructuring of ILs characteristic of cations with longer aliphatic chains as well as the ILs liquid-vapour interfacial structure. The CG nature of these models enables the simulation of very long time scales, which are needed to computer reliable results of dynamic and interfacial properties.  For  [C$_{>4}$MIM] [BF$_4$] the break in symmetry associated to the liquid-vapor interface induces nanostructuring in polar and non-polar domains in the direction perpendicular to the interface plane, hence mimicking the behavior observed in the bulk phases.
\end{abstract}

\section{Introduction}

Ionic Liquids (ILs) bring new opportunities in synthesis and catalysis~\cite{welton} and energy storage~\cite{chmiola}. Recently, ILs have been
considered in applications concerned with 
tunable lubrication, a new field called {\it tribotronics}\cite{GLAVATSKIH2008934,C6CP04405K,li2014}. Tunability is a desirable property of new lubricants, which can be achieved by modifying the structural and dynamic 
properties of  ILs 
using external fields. This idea might provide an approach  to 
modulate friction forces, and possibly to achieve superlubricity states, which still represents nowadays   
the ``holy grail'' of nanolubrication research~\cite{urbakh2004,RevModPhys.85.529}. Achieving superlubricant states on demand will enable significant reduction of energy dissipation and, therefore, the enhancement of the performance of many devices in 
technology and industry. 

Experimental studies have uncovered the importance of confinement  at the nanoscale in determining the lubrication response of ILs. 
When ILs are confined between negatively charged surfaces, the application of pressure induces the thinning of the film, by expelling pairs of ions layers, hence preserving the electroneutrality of the confined region~\cite{perkin}. Furthermore, the friction force depends on the composition of the ionic liquid, and its degree of 
hydrophilicity and 
affinity to adsorb water~\cite{espinosa2014ionic,smith2014molecular}. This makes necessary to expand the range of theoretical tools, to investigate 
the variability of composition that have been considered so far in experiments.

Recently, we explored 
the physics of tunable lubrication using a range of coarse grained models incorporating different levels of complexity~\cite{fajardo-scirep,fajardo-jphysclett}. We addressed the impact of substrate structure and water adsorption on friction. 
Coarse grained (CG) models offer advantages to address these problems, because the 
dynamics of ionic liquids is very slow, with diffusive times varying between 1--100~ns, and the dynamic response might decrease further with respect to
bulk conditions, in
nano-confinement. CG models can be used to
perform simulations over
 long time scales, 10--100~ns, at a considerably lower 
computational cost than atomistic models. These long time scales  
ensure a proper sampling of the liquid properties. However, 
most of the studies on friction mentioned above, focused on specific ILs, the butylimidazolium family, and hence the role of the length of the
hydrocarbon chain of the cation or ILs mixtures has not been addressed. 
The length of the aliphatic chain of the cation is 
of particular interest in this context, since  it has been demonstrated using 
computer simulations that 
chains longer than four carbon atoms induce the formation of heterogeneous nanometer scale structures~\cite{CanongiaLopes2006}. The structuring of these 
``mesoscopic'' liquids 
is associated to the unfavourable interactions of 
non-polar and charged groups. 
Such interactions induce the 
formation of globular and lamellar structures, which are reminiscent of the ones observed in {\it e.g.} surfactants and lipids. Being able to model these structures in the
context of non-equilibrium conditions arising from 
friction processes is, therefore, of considerable interest.

In this work we perform molecular dynamics simulations of  the [C$_\mathrm{n}$MIM][BF$_4$], with  $\mathrm{n}=2$,$4$,$6$,$8$ family of ILs, using a new set of CG models. We extend previous models to investigate cations featuring longer alkyl chains. We focus on the impact that the 
length of the aliphatic chain of the cation has on the structural, interfacial and dynamic properties. This work is a starting 
point for CG-simulations of 
electrotunable friction with 
ILs, which will be presented in a paper to follow.


\section{Model and Methods}
\begin{figure}[!tb]
  \centering
    \includegraphics[scale=1.5]{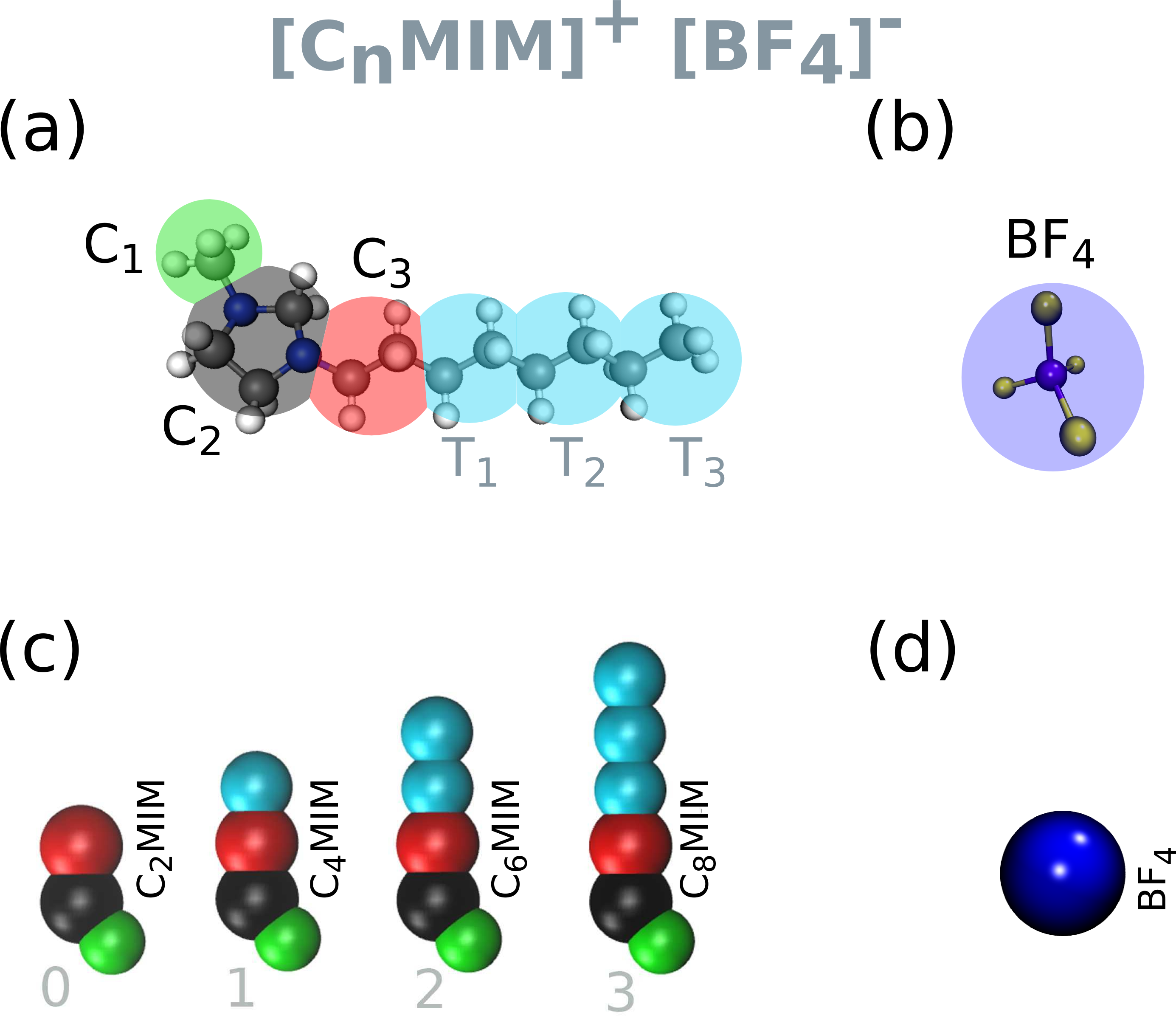}
      \caption{(a) and (b) Mapping approach used to model the ILs using the CG model. Sketch of the (c) cations and (d) anion investigated in this work.
      } \label{Fig0}
\end{figure}

The model considered here builds on the force-fields introduced in previous works~\cite{Roy2010,Merlet2012}. These force-fields define the imidazolium and alkyl groups as pseudoatoms. We have extended this CG model to investigate 
the thermophysical properties of  1{-}n{-}3{-}methyl-imidazolium tetrafluoroborate [C$_\mathrm{n}$MIM][BF$_4$] ILs, with  $\mathrm{n}=2$,$4$,$6$,$8$ (see Figure~\ref{Fig0} for an illustration of the chemical structures and the mapping approach used for the CG model). 
The 
cations consist of a charged head group (polar group), modelled using  three interaction sites, C$_1$, C$_2$ and C$_3$,
and nonpolar alkyl side-chain groups, T$_i$, with $i = 1, 2, 3$, 
which we  model using  
a two-to-one mapping scheme, namely 
two 
methyl groups, CH$_2$, in the aliphatic chain of the  full atomistic system are represented by
one pseudoatom (see 
Figure~\ref{Fig0}). 
The first member of the family, [C$_2$MIM][BF$_4$], does not have T$_i$ pseudoatoms. 
ILs with T$_{1-3}$ represent 
[C$_4$MIM][BF$_4$] 
[C$_6$MIM][BF$_4$] 
and [C$_8$MIM][BF$_4$], respectively. We use the subscript ``$c$'' (T$_c$) or ``$e$'' (T$_e$) to indicate the nonpolar group at the center or at the edge of the cation molecule, respectively.

The pseudoatoms interact via 
dispersion and coulombic interactions: 
\begin{equation}
U_{ij}(r_{ij})=4\varepsilon_{ij} \left[ \left( \frac{\sigma_{ij}}{r_{ij}} \right)^{12} - \left( \frac{\sigma_{ij}}{r_{ij}} \right)^{6}\right] + \frac{q_iq_j}{4\pi\varepsilon_0 r_{ij}}
\label{Pot}
\end{equation}
The dispersion interactions were modelled using the Lennard{-}Jones (LJ){-}potential, with effective
diameter, $\sigma_{ij}$,  and interaction strength, $\varepsilon_{ij}$. 
The cross interactions between pseudoatoms are obtained using standard combination rules, $\sigma_{ij} = (\sigma_{i} + \sigma_{j})/2$ and $\varepsilon_{ij} = \sqrt{\varepsilon_{i} \varepsilon_{j}}$. We compile in Table~1 all the parameters for the different pseudoatom types, including both
intermolecular and intramolecular contributions. 
Following previous strategies we 
used 
fractional charges to describe the ion charge, in order to take  
into account  
effective polarization effects. 

\begin{table}[!ht]
 \caption{Force Field parameters for [C$_\mathrm{n}$MIM][BF$_4$] IL{-}family.  $^\mathrm{a}$ For [C$_2$MIM][BF$_4$] the mass is 29.07~g~mol$^{-1}$.}
 \centering 
 \begin{tabular}{|c|c|c|c|c|} 
  \hline\hline
  Site &  M / (g mol$^{-1}$) & $\sigma_{i,i}$ / (nm) & $\varepsilon_{i,i}$ / (kJ mol$^{-1})$ & $q_i$(e)\\  
  \hline
  C$_1$ &  15.04 & 0.3410 & 0.360 & 0.1888 \\ 
  C$_2$ &  67.07 & 0.4380 & 2.560 & 0.3591 \\\
  C$_3^\mathrm{a}$ & 28.06 & 0.4380 & 1.240 & 0.2321 \\ 
  T$_\mathrm{c}$ &  28.06 & 0.3833 & 1.240 & 0.0000 \\
  T$_\mathrm{e}$ &  29.07 & 0.3833 & 1.240 & 0.0000 \\ 
  BF$_4$ & 86.81     & 0.4510   & 3.240 & -0.7800 \\ 
  \hline
 \end{tabular}
\end{table}

We  report in the next section data for the bulk density ($\rho$), radial distribution functions (RDFs), diffusion coefficients (D), viscosity ($\eta$) and surface tension ($\gamma$) as a function of the chain length of the cation. 
The simulations of liquid-vapor interfaces were performed with 
orthorhombic simulation cells, $L_x \sim  L_y < L_z$, and a liquid slab with a film thickness between 11 and 30 nm for [C$_2$MIM][BF$_4$] and [C$_8$MIM][BF$_4$] ILs, respectively. The cross sectional area varied between 6 and 13 nm, depending on the system size (between 1600 and 9600 molecules).   
The slabs were surrounded by vacuum regions, with a thickness of 3 times the slab width.  

The computations of the density and diffusion coefficients were performed using simulation boxes containing 468 ion pairs. 
The system was equilibrated for 20~ns at constant pressure and temperature using the Berendsen barostat~\cite{berendsen} and the v-rescale thermostat \cite{vrescale} using coupling constants of 1~ps and 0.2~ps, respectively. 
The equilibrated configurations were used to perform additional simulations in the NVT ensemble, using the 
v-rescale algorithm with a time coupling constant of 0.2~ps.
The production phase spanned $20$ ns for the short chains, 
[C$_2$MIM][BF$_4$], 
and up to $0.4$ $\mu$s for the cations with longer chains. We found that this simulation time is necessary to reach the diffusive regime and to obtain reliable 
diffusion coefficients. These simulations are
significantly longer than the ones reported in several simulation studies, particularly using atomistic models.~\cite{tsuzuki2009,voroshylova2014atomistic}
The trajectories were generated using 
a time step of either 2 or 10~fs. Advancing the discussion below, we found that the longer time step produces stable trajectories (see Figure~5 for a comparison of mean square displacements obtained using both time steps), in line with previous simulations of biomolecules using  CG models, {\it e.g.} the MARTINI force-field. 
The long range electrostatic interactions were computed using the 3D Particle Mesh Ewald (PME)~\cite{pme} summation. 
In line with previous reports, we found that the density and  particularly the surface tension depend on the cut-off used to truncate the LJ interactions~\cite{doi:10.1063/1.480192}. 
The range of the dispersion interactions is an important variable in the definition of the force-field.  We chose to perform all the computations using the full LJ potential, by employing the PME summation method. This approach 
ensures that the force-field  gives the same results 
using both Molecular Dynamics and Monte Carlo simulations.

The intramolecular bonds between a pair of pseudoatoms were modelled using rigid constraints and the LINCS algorithm \cite{LINCS}. Three body interactions were modelled using a harmonic potential (see Table 2). The angle between 
three consecutive pseudoatoms in the aliphatic chain was set to 180$^o$, {\it i.e.} similar to the equilibrium angle used in simulations of aliphatic chains in lipids~\cite{doi:10.1021/jp036508g}. The electrostatic interactions were computed in full using the Particle Mesh Ewald method, and
the dielectric permittivity was set to 1, because the screening of electrostatic interactions  by the electronic polarizability of the ions is taken into account already with the reduced values of the
effective charges. All the trajectories were generated with GROMACS v. 5.1.4\cite{gromacs}.


\begin{table}[!ht]
 \caption{Parameters used to model the intramolecular interactions of the cation, $\mathrm{C_n MIM}$.  The three body interactions are modelled using a harmonic potential $V(\theta)=\frac{1}{2} k_\theta (\theta - \theta_0)^2$.}
 \centering 
 \begin{tabular}{ |c|c| c|c| c|} 
  \hline\hline
  Interaction  &  $r_b$ (nm) & Interaction & $\theta_e$ (deg) & $k_b$ (kJ/(mol rad$^2$))    \\ 
  \hline
   C$_1$-C$_2$          & 0.2700   & C$_1$-C$_2$-C$_3$ &   130 & 500     \\ 
   C$_2$-C$_3$          & 0.3025   & C$_2$-C$_3$-T$_\mathrm{c}$ &   180 & 125     \\
   C$_3$-T$_\mathrm{c}$          & 0.3025  & C$_3$-T$_\mathrm{c}$-T$_\mathrm{c}$ &    180 & 125      \\
   T$_\mathrm{c}$-T$_\mathrm{c}$          & 0.3025  & T$_\mathrm{c}$-T$_\mathrm{c}$-T$_\mathrm{e}$ &  180   & 125         \\
   T$_\mathrm{c}$-T$_\mathrm{e}$          & 0.3025   &                &           &                \\
  \hline
  \end{tabular}
\end{table}


Some bulk properties, such as density and radial distribution function, obtained with the CG model were compared with those computed using the atomistic force-fields of Canongia-Lopes and P\'{a}dua~\cite{CanongiaLopes2006}. To this end, we performed 5~ns NPT simulations of 512 ion pairs of [C$_\mathrm{n}$MIM][BF$_4$] IL{-}family at 1~bar and for different temperatures.

We performed additional simulations to compute the viscosity of the ILs at selected temperatures. Our approach builds on non-equilibrium molecular dynamics (NEMD) simulations via the periodic perturbation method~\cite{hess}. The method relies on the generation of a velocity field, $v_x(z)$, which is imposed along one direction of the simulation box ($z$ in our case). The field satisfies the Navier-Stokes equation at the stationary state,
\begin{equation}
\label{navstok}
a_x(z) = -\frac{\eta}{\rho} \frac{\partial^2 v_x(z)}{\partial z^2}
\end{equation}

\noindent where $\rho$ is the average density of the fluid, $a_x(z)$ is the  acceleration (external force) that drives the velocity field. The acceleration changes along the $z$ axis in the box frame of reference according to,

\begin{equation}
\label{velfield}
a_x(z)=A \cos(kz)
\end{equation}
where $A$ is the maximum amplitude of the acceleration and the vector $k=2\pi /L_z$ is defined by the length of the box in the $z$ direction, $L_z$. The corresponding velocity is given by, 
\begin{equation}
\label{velp2}
v_x(z) = V \cos (kz)
\end{equation}

\noindent with $V = A \rho/\eta k^2$.  The viscosity is given by, 

\begin{equation}
\eta = \frac{A}{V} \rho k^2.
\end{equation}

\noindent $V$ is computed in the simulation from the instantaneous velocity profile. We explored different accelerations,  and the results reported below were obtained with 5~m~s$^{-2}$, and prismatic boxes with dimensions \{$L_x, L_y, L_z$\} = \{1,1,3\}.  This set up provides precise estimates of the viscosity. A typical simulation box consisted of 2400 molecules, and averages were obtained over trajectories spanning 10~ns. We performed these simulations with the shorter time step, 2~fs. The performance of the NEMD method against the equilibrium Green-Kubo (GK) approach has been tested before in simulations of
non-polar and polar fluids~\cite{hess,ZHENG20191}, showing that the GK 
and NEMD approaches predict the same results within the uncertainty 
of the computations. 

\section{Results}

\subsection{Thermodynamic and structural properties}

\begin{figure}[!ht]
  \centering
    \includegraphics[scale=0.65]{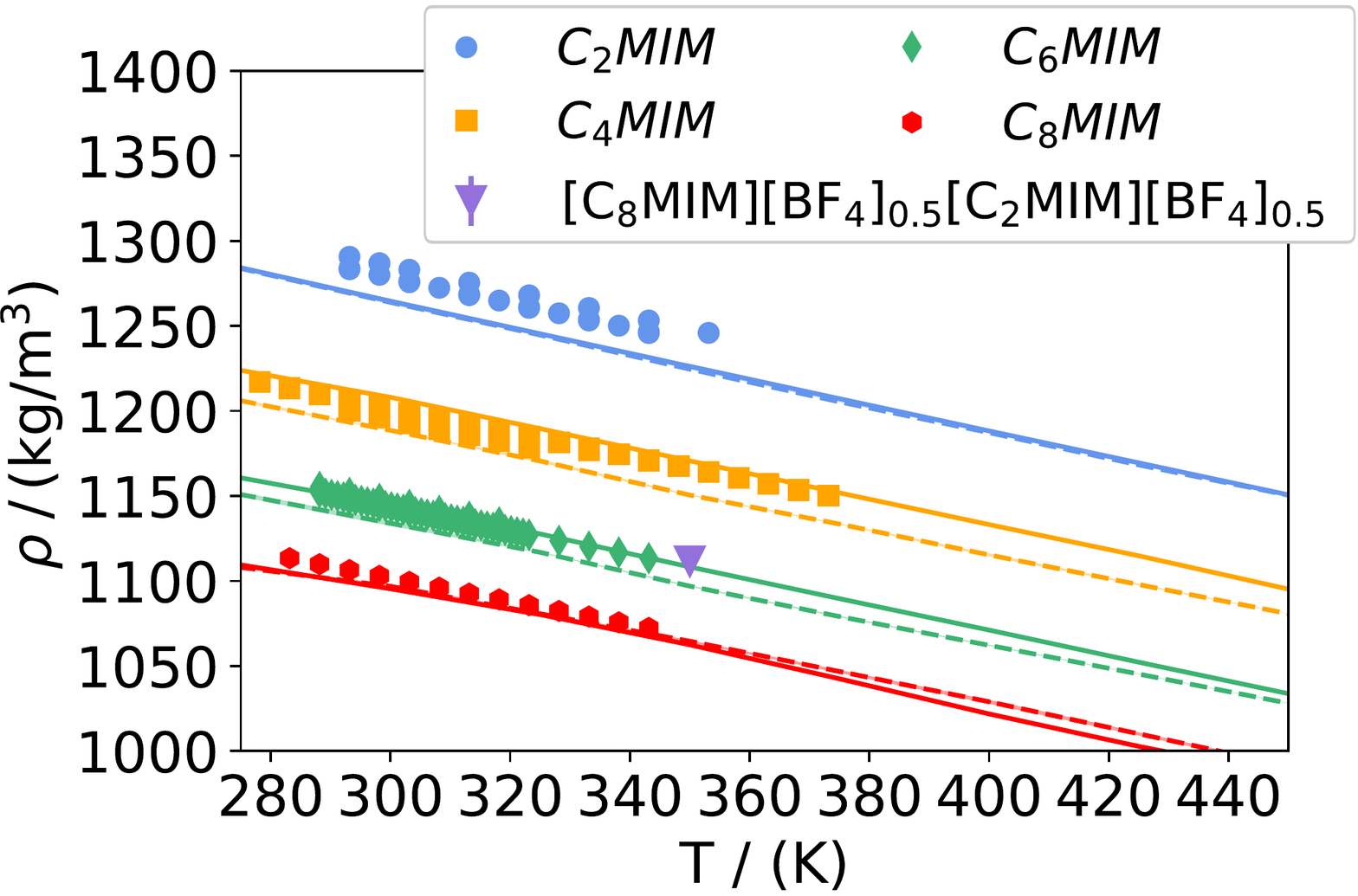}
      \caption{Temperature dependence of the density of the [C$_\mathrm{n}$MIM][BF$_4$] family of ILs at $\sim$1~bar. The symbols represents the experimental results for  [C$_2$MIM][BF$_4$]~\cite{Song2014JChemEngD,Ciocirlan2014JChemEngD,VERCHER2015174}, [C$_4$MIM][BF$_4$]~\cite{Chaudhary2014,SALGADO2014101,Wu2015,SRINIVASAKRISHNA2015350}, [C$_6$MIM][BF$_4$]~\cite{Song2014JChemEngD,PAL2012157,SANMAMED200796}, [C$_8$MIM][BF$_4$]~\cite{Kumar2008}. The continuous lines indicate the results from the CG simulations and the dashed lines the density obtained using the OPLS-AA force-field~\cite{CanongiaLopes2006}. The triangle-down represents the computed density of the binary mixture [C$_8$MIM][BF$_4$]$_{0.5}$--[C$_2$MIM][BF$_4$]$_{0.5}$ modelled with the CG force-field. 
      }
    \label{eos}
\end{figure}

The performance of the force-fields is commonly assessed by computing the equation of state. We have analyzed this as a function of the length of the aliphatic chain of the cation and the temperature. Analyses of experimental data show that at a given temperature, the density of the IL decreases with the size of the C$_\mathrm{n}$MIM cation. We show in Figure~\ref{eos} that the CG model reproduces this  behavior quantitatively. The predicted densities are in good agreement with the experimental data, with deviations not exceeding 1.5\%, being 1.4\% for C$_2$MIM, 0.45\% for C$_4$MIM, 0.11\% for C$_6$MIM, 0.60\% for C$_8$MIM. The slope of the density and temperature curves agrees closely with the experiment too (see Figure~\ref{eos}) indicating that the CG models reproduce the thermal expansion of the ILs. Overall, the performance of the CG model is in line
with data reported before with atomistic and other CG models\cite{liu2004,liu2006,koddermann2007,sambasivarao2009,liu2012,chaban2011}. 

In addition to pure fluids, we have performed simulations of one equimolar binary mixture of ILs, consisting of cations of different chain length, 
[C$_4$MIM][BF$_4$]$_{0.5}$--[C$_8$MIM][BF$_4$]$_{0.5}$ IL. 
The predicted density, 1111.7$\pm 0.2$ kg~m$^{-3}$,  is approximately between the average densities of the pure components (1115.5$\pm 0.2$ kg~m$^{-3}$), indicating that the mixing is nearly ideal. This observation agrees with the experimental measurements of  [C$_\mathrm{n}$MIM][BF$_4$] ILs mixtures, which reported small excess volumes upon mixing\cite{Song2014JChemEngD}.   


\begin{figure}[!ht]
  \centering
  \begin{tabular}{cc}
  \includegraphics[scale=0.44]{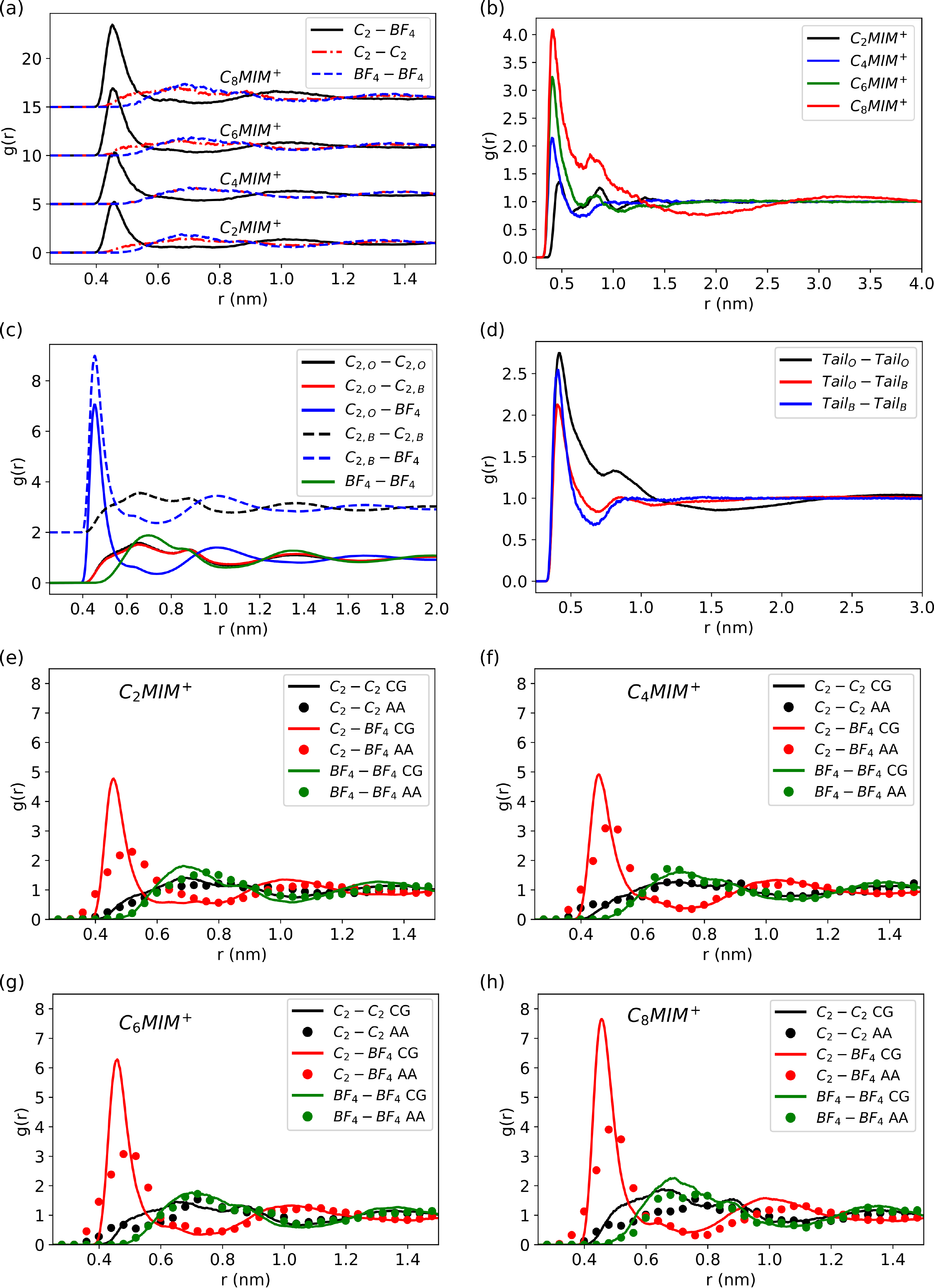}
      \end{tabular}
      \caption{RDF for the [C$_\mathrm{n}$MIM][BF$_4$] family of ILs, $\mathrm{n}=2,..,8$. (a) RDF between the imidazolium groups (using the C$_2$ bead, see Figure~\ref{Fig0}) and the BF$_4$ pseudoatom.  (b) RDF between the last bead, $\mathrm{T}_\mathrm{e}$, in the tails of the aliphatic chains of the cations. (c) RDF of the mixture [C$_\mathrm{n}$MIM][BF$_4$]$_{x}$--[C$_\mathrm{m}$MIM][BF$_4$]$_{1-x}$ IL , $x=0.5$. The sub-indices $O$ and $B$ refer to octyl and butyl cations, respectively. (d) RDF between the charged groups and tails. (e-h) Comparison of RDFs obtained with the CG model presented here and the OPLS/AA force-field introduced in Ref.~\cite{CanongiaLopes2006} for [C$_\mathrm{n}$MIM][BF$_4$] as indicated in each panels. All the data correspond to 350~K and 1~bar.
       }
    \label{rdf}
\end{figure}

To gain microscopic insights into the spatial organization of the IL molecules in  bulk, we computed the radial distribution functions (see Figure~\ref{rdf}).  We note that the model employed here for C$_4$MIM is inspired by previous works\cite{Roy2010}, but it differs in the number of beads employed, which we have expanded here by one bead in order to build a unique force-field for the series of C$_\mathrm{n}$MIM ILs. The RDFs agree very well with the previous  CG model. The comparison with the atomistic models also shows that this force-field well captures both the size of cations and anions as well as the characteristic decay of electrostatic correlations, as shown by the damped oscillations  (see Figure~\ref{rdf}(e-h)).  The RDFs for cations of different size (see Figure~\ref{rdf}(e)) feature the characteristic out of phase oscillations  for cation-cation and cation-anion correlations, even for the longer aliphatic chain we have simulated, C$_8$MIM. This oscillatory decay is consistent with the well known general behavior observed in molten salts~\cite{hansen}.

\begin{figure}[!ht]
  \centering
        \includegraphics[angle=-90, scale=0.6,trim=6cm 7cm 6cm 7cm]{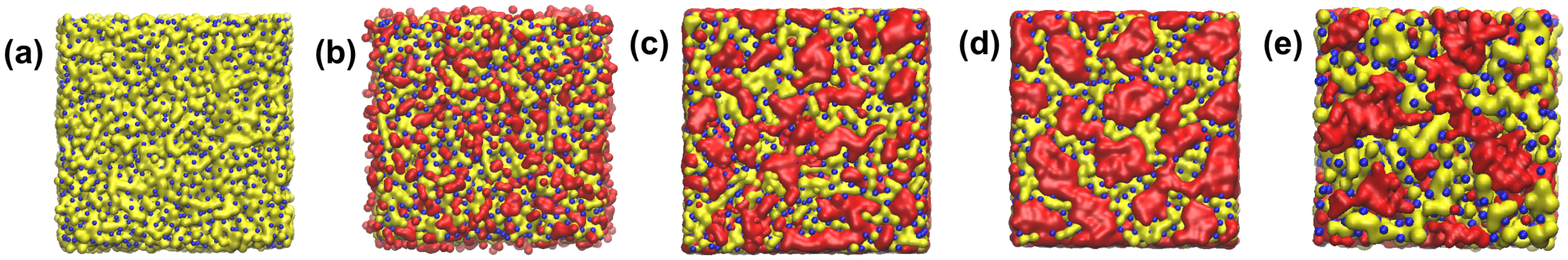} 
     \\[3pt]
\begin{subfigure}[t]{0.03\textwidth}
    \textbf{(f)}
  \end{subfigure}
    \begin{subfigure}[t]{0.45\textwidth}
    \includegraphics[width=\linewidth,valign=t]{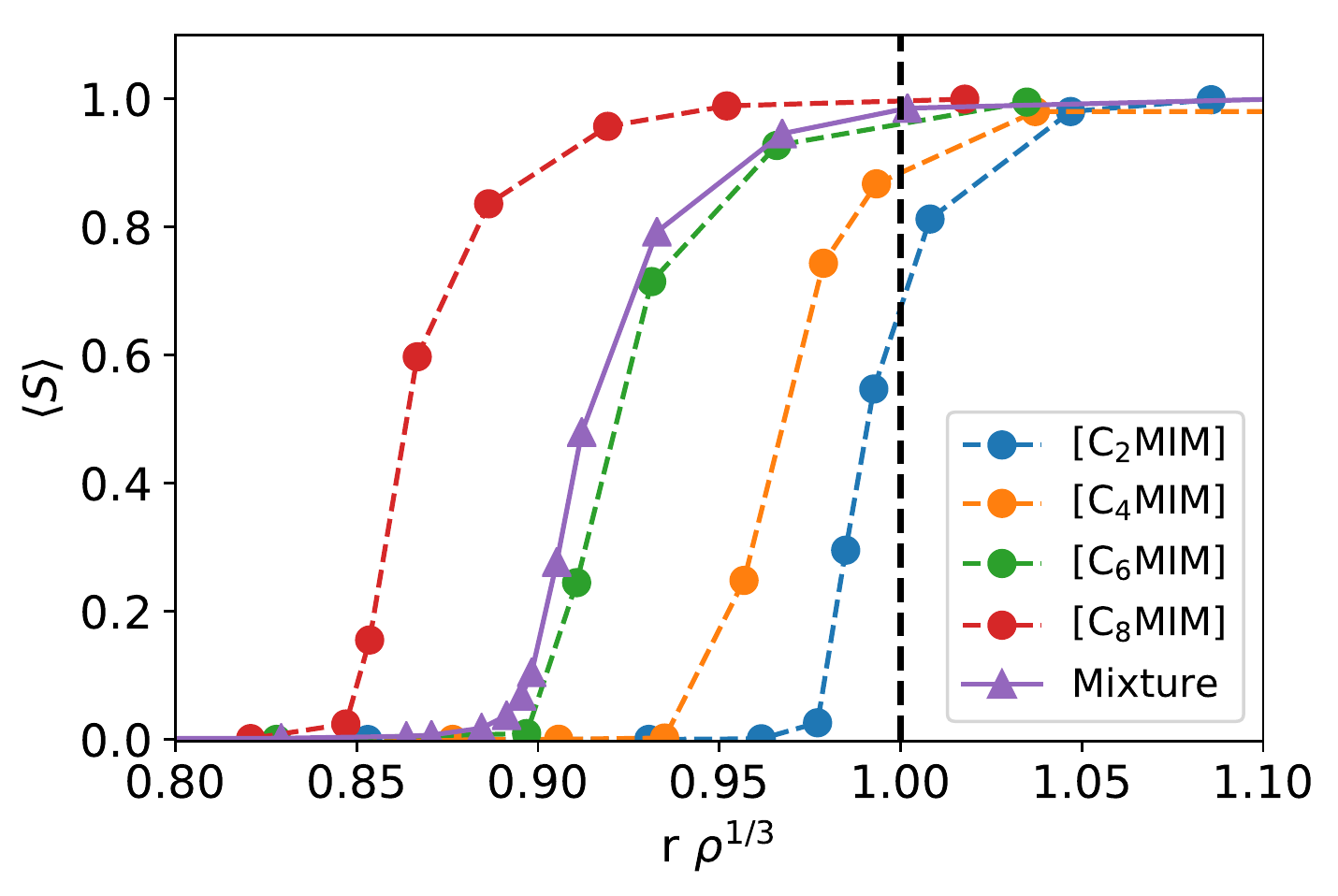}
  \end{subfigure}
      \caption{(a)-(e) Snapshots of representative configurations of ILs,  [C$_{n}$MIM][BF$_4$] highlighting the formation of nanodomains. All the simulations were performed at 
	  $350$~K and 1 bar . (a)  [C$_{2}$MIM][BF$_4$] (b) [C$_{4}$MIM][BF$_4$] (c) [C$_{6}$MIM][BF$_4$]  and (d) [C$_{8}$MIM][BF$_4$]. The panel (e) represents the nanodomains for the mixture, [C$_4$MIM]$_{0.5}$[BF$_4$]--[C$_8$MIM]$_{0.5}$[BF$_4$]. The  surfaces in yellow represent the polar region of the cation (see Figure 1). The regions coloured in red represent the aliphatic chains, and the spheres colored in blue represent the anions. The box size for the this panel is  
	  7~nm, about half the box length for panels (a-d). (f) Average cluster size as a function of the cut-off distance, $r$, employed to construct the clusters, reduced  by the average distance between the anions, $\rho^{-1/3}$. The vertical bar indicates the average distance between anions for a non interacting system.}
    \label{snapshots}
\end{figure}

The chain length of the cation has a considerable impact on the spatial organization of the non-polar groups (see Figure~\ref{rdf}(a)). The RDFs of C$_{2,4}$MIM ILs decays to the ideal value ($g(r)=1$) within $\sim$ 1~nm. The emergence of homogeneous structures in the liquid is already observed in C$_6$MIM, which converges towards the ideal gas structure  at $\sim 1.5$~nm, while the inhomogeneity of the fluid becomes dominant for  C$_8$MIM ILs, which features long wave length oscillations, (typically $\sim$3~nm) indicating the formation of nanodomains.  The oscillations arise from the aggregation of the aliphatic chains with each other (see  Figure~\ref{snapshots}, top panels), which leads to interpenetrated non-polar and charged nanoscopic regions (see panel (d) in Figure~\ref{snapshots}-top panel). 

We show in Figure~\ref{snapshots} how the structure of the bulk liquid coarsens with increasing cation size. To highlight the non-polar and charged domains we have represented the aliphatic chains and charged regions as surfaces, using the VMD software~\cite{HUMP96}. These surfaces were constructed by rolling a spherical probe of radius 0.15~nm around the charged and non-polar regions,
separately.  The resulting surfaces are represented with different colors, while the anions are represented as spheres (see Figure~\ref{snapshots}). While the structure of [C$_2$MIM][BF$_4$] IL is homogeneous, our CG force-field predicts the formation of  polar/non-polar rich regions with increasing cations size. In all cases the anions accumulate around the charged part of the cations. The CG model displays a continuous transition from a homogeneous structure 
for short 
alkyl{-}chains to a heterogeneous structure featuring nanodomains, which is favored
for long alkyl-chains. The observation of nanodomains  is consistent with previous simulations using atomistic and 
CG models\cite{wang2005,CanongiaLopes2006}.  

We have extended our analysis and simulated the IL mixture (see Figure~\ref{snapshots}(e)). This snapshot also shows the formation of extended polar/non-polar regions. Advancing the discussion below, the clustering structure of the mixture lies between that of C$_4$MIM and C$_8$MIM. 
Note that the simulation box size for the mixture is $\sim 7$~nm {vs.}
12-13~nm, twice as large, for the pure ILs fluids. This size different accounts for the different size of the domains relative to the box size.

To gain insights into the heterogeneous structures formed in the ILs,
we calculated the average cluster size (ACS), which is defined as $\left< S \right> = \frac{1}{N}\frac{\sum_s s^2 n_s}{\sum_s  s n_s}$, where the sums run over cluster of size, $s \ge  1$, with $n_s$ being the number of clusters of size $s$, and $N$  the total number of particles used to construct the clusters. The definition $\left < S \right >$  varies in the interval $\left [1/N,1 \right]$. We calculated the ACS  
using a distance criterion that was applied to the anions, BF$_4$ ``only''. As the latter accumulate in the polar region, the clustering analysis of the anions provides a good reference to quantify the heterogeneity of the liquid. Hence, whenever two BF$_4$ anions were at a distance closer than a given cut-off, $r_c$, they  were deemed to be part of the same cluster.

We show in Figure~4-bottom panel the ACS distribution as a function of the cut-off distance, $r_c$. 
The distribution 
features the characteristic sigmoid behavior with $r_c$, with an inflection point indicating the approximate inter-anion distance at which the cluster transitions from a finite size aggregate to a percolating cluster, which spans the whole simulation box. The transition for the shorter aliphatic chains, C$_2$MIM appears approximately at the inter-ionic distance corresponding to an uncorrelated system, $\rho^{-1/3}$, {\it i.e.} the distance is determined by the number of ions. As the length of the size of cation increases the transition in $\left< S \right >$ shifts to distances shorter than $\rho^{-1/3}$. This is the expected behavior for a fluid where the anions are not distributed homogeneously, and therefore it is indicative of aggregation.  The accumulation of the anions in the charged regions is favored at longer aliphatic chains, and this is reflected in the systematic shift of the inflection point of $\left < S \right >$ to shorter distances (see in Figure~\ref{snapshots}). 
We have computed the ACS distribution of the mixture [C$_4$MIM][BF$_4$]$_{0.5}$--[C$_8$MIM][BF$_4$]$_{0.5}$ BF$_4$ (see Figure 4-e). As expected the inflection point in  $\left< S \right>$,  shows the appears at distances shorter than that of the uncorrelated system (see Figure 4-e). 
The transition in $\left< S \right>$ appears at the same distance as that of C$_6$MIM, indicating that the suggesting the clustering structure of the mixture and the pure C$_6$MIM and [C$_4$MIM][BF$_4$]$_{0.5}$--[C$_8$MIM][BF$_4$]$_{0.5}$ are very similar. This behavior mimics our results for the equation of state, where the density of the mixture was found to be the same as that of C$_6$MIM (see Figure~2).

\subsection{Dynamic properties}
Experimental studies have shown that the self-diffusion coefficient of the cations and anions decreases as the size of the cations increases~\cite{tokuda2005}. A connection between the observed dependence of the dynamics and  the aggregation behavior of the ions was noted before~\cite{CanongiaLopes2006}. 

The prediction of the self-diffusion coefficient with classical force-fields has proven challenging. Earlier studies showed that atomistic force-fields predict too low diffusion coefficients, as compared with the results obtained from NMR measurements~\cite{tsuzuki2009,C4CP05550K}. This issue has been addressed by including polarization effects in the force-field or reparametrizing the parameters defining the dispersion 
interactions with the LJ-potential. These approaches lead to better  agreement with the experimental results~\cite{borodin,C4CP05550K,tsuzuki2009,koddermann2007}. 
However, a systematic analysis assessing the impact of polarization alone is not always trivial,  since polarization  introduces additional changes in the functional form of the force-field.

\begin{figure}[tb]
  \centering
  \begin{subfigure}[t]{0.03\textwidth}
    \textbf{(a)}
  \end{subfigure}
   \begin{subfigure}[t]{0.45\textwidth}
    \includegraphics[width=\linewidth,valign=t]{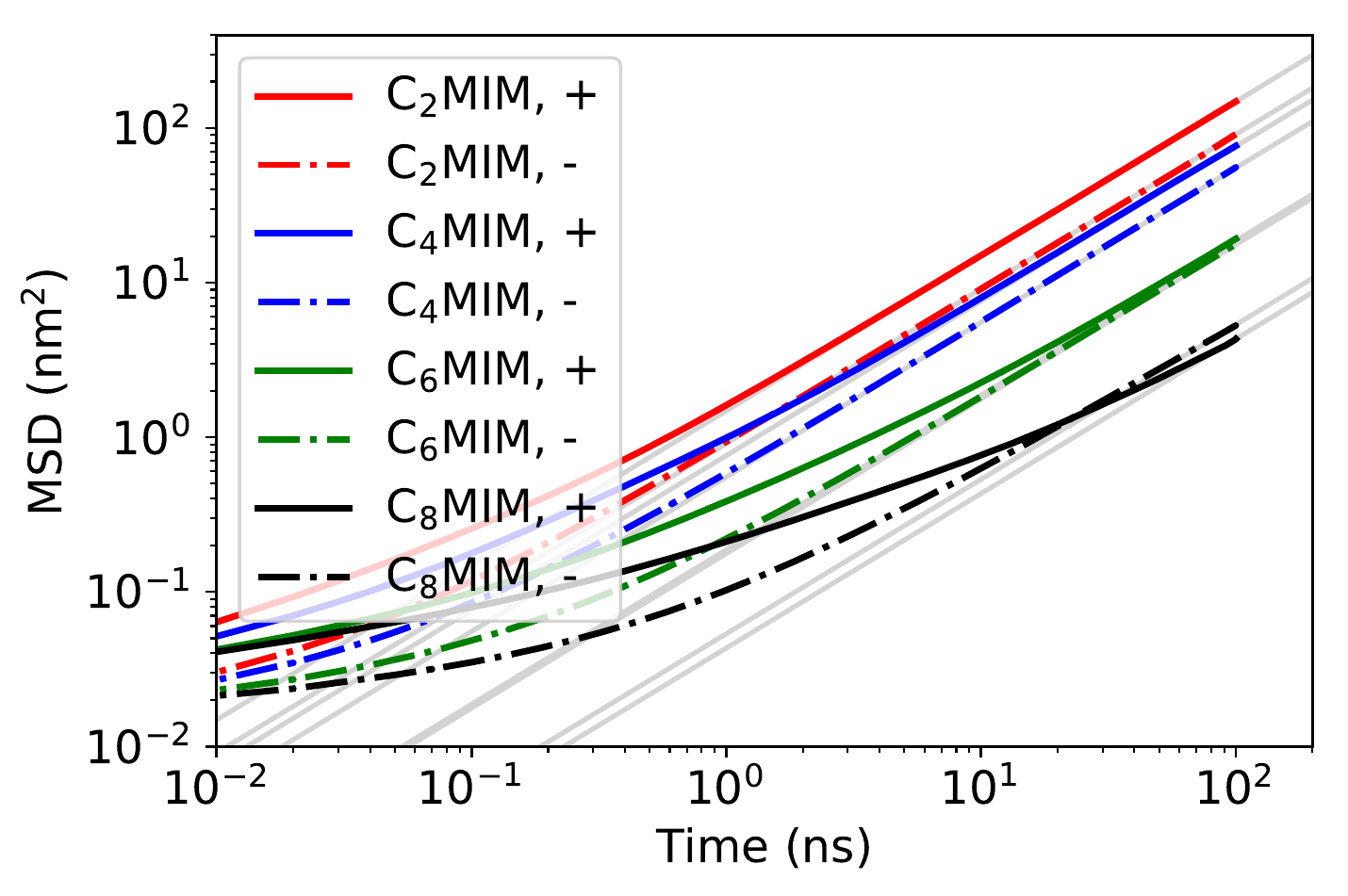}
  \end{subfigure}
  \begin{subfigure}[t]{0.03\textwidth}
    \textbf{(b)}
  \end{subfigure}
    \begin{subfigure}[t]{0.45\textwidth}
    \includegraphics[width=\linewidth,valign=t]{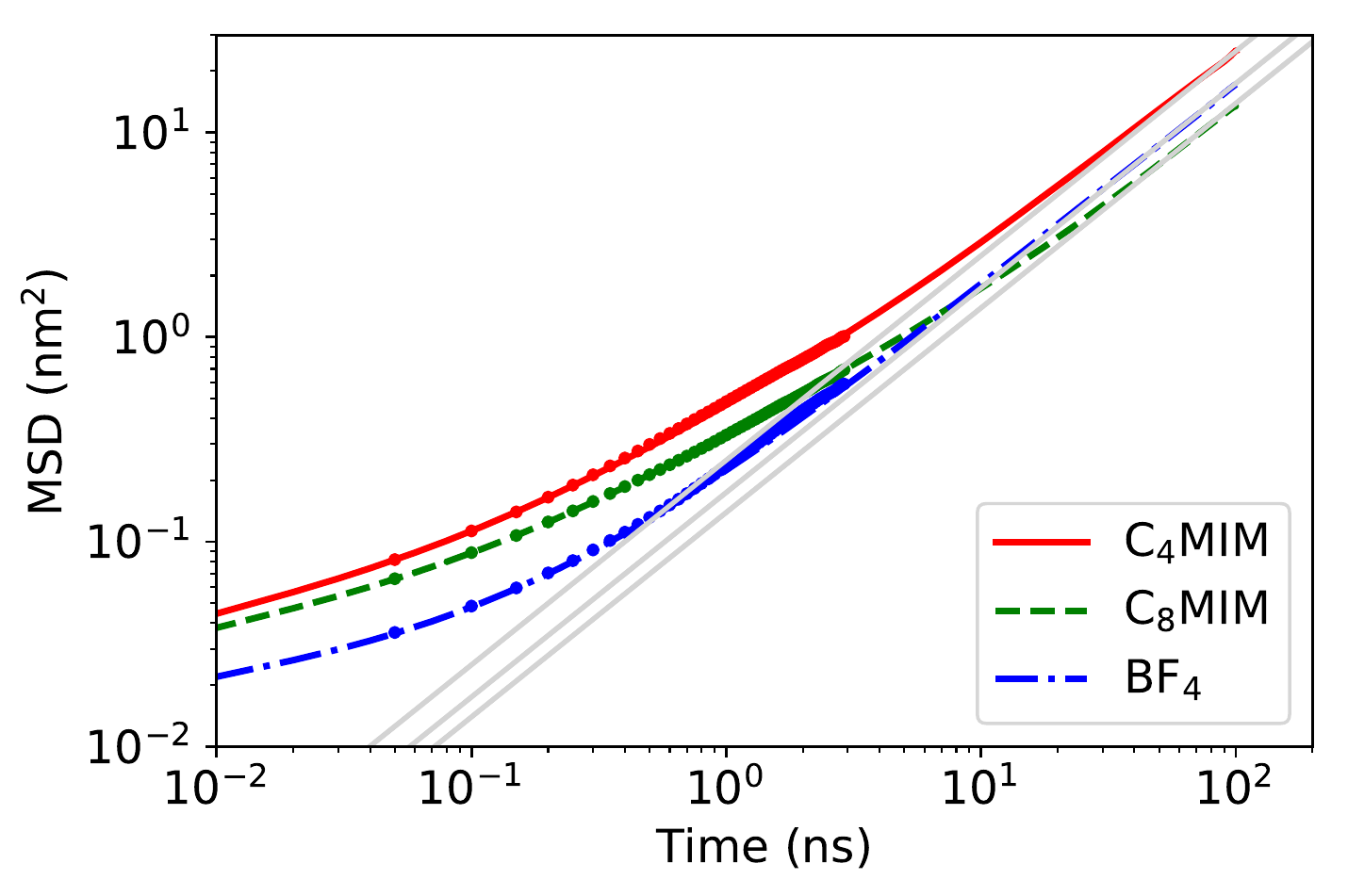}
  \end{subfigure}
    \\[3pt]
    \begin{subfigure}[t]{0.03\textwidth}
    \textbf{(c)}
  \end{subfigure}
    \begin{subfigure}[t]{0.45\textwidth}
    \includegraphics[width=\linewidth,valign=t]{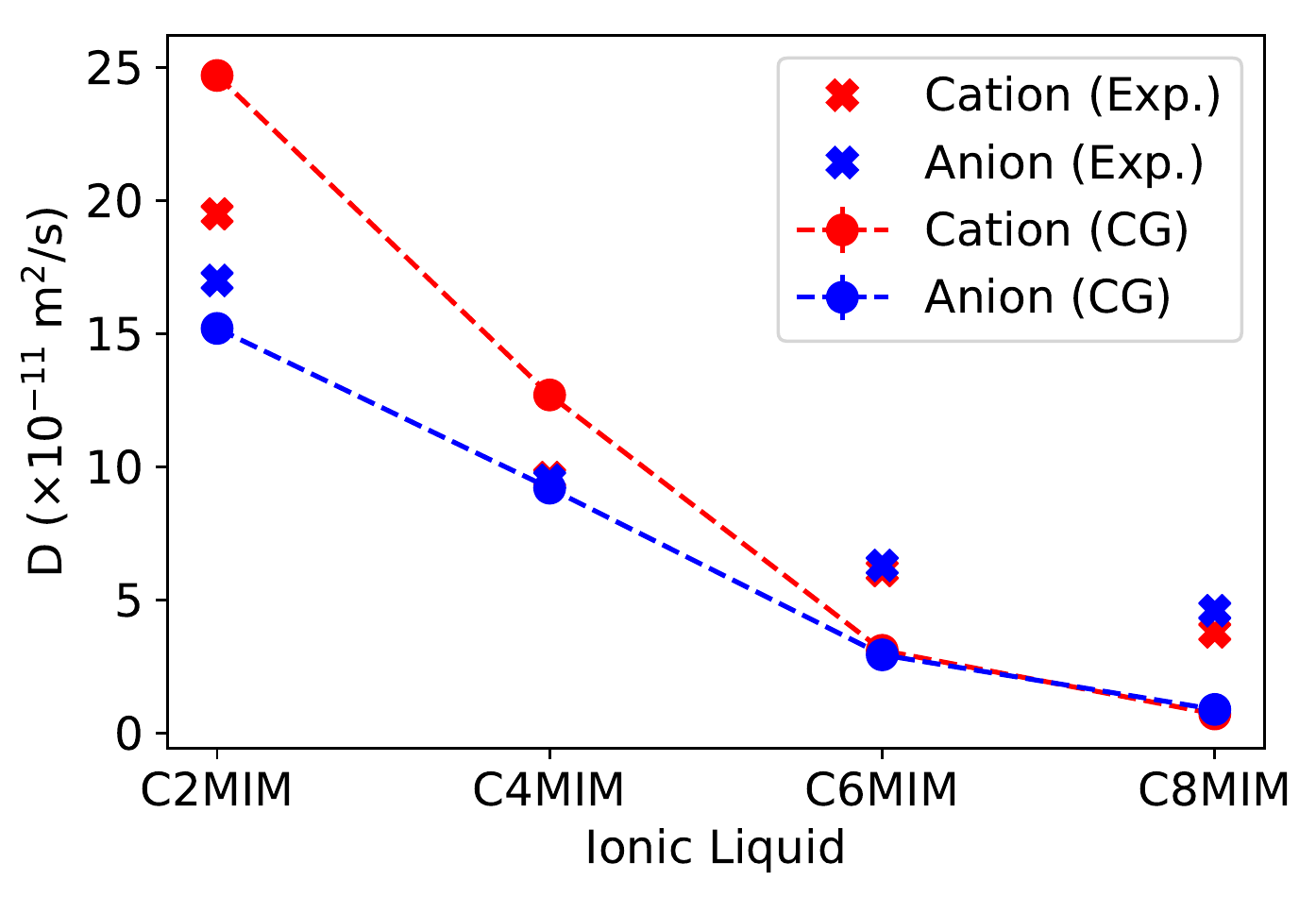}
  \end{subfigure}
      \caption{(a) Time  dependence of the mean square displacement (MSD) for [C$_\mathrm{n}$MIM][BF$_4$] ILS ($\mathrm{n}=2,4,6,8$). The oblique lines are a guide to the eye, and they indicate the slope of the MSD corresponding to the diffusive regime, $t^{1}$. The $+$ and $-$ signs in the labels indicate results for cations and anions, respectively. (b) MSD for the mixture [C$_\mathrm{n}$MIM][BF$_4$]$_{0.5}$--[C$_m$MIM][BF$_4$]$_{0.5}$ BF$_4$. The points at times $<3$~ns were obtained using a time step of 2~fs. All the data were obtained at 350 K and 1 bar. (c) Diffusion coefficients of cations and anions as a function of the size of the cation at 350 K and 1 bar. Numerical data and references for the experimental data are give in Table~3.  All the data were obtained at 350 K and 1 bar. The experimental data are taken from the references given in Table~3. 
 }
    \label{fig:diffusion}
\end{figure}

We assess here the accuracy of the CG models by computing the 
self-diffusion coefficient of the cations and anions. The diffusion coefficient, $D$, was computed using  Einstein's equation and the mean square displacement (MSD): 
\begin{equation}
D = \lim_{t \to \infty} \frac{ \langle \lvert  {\bf r}_i(t) - {\bf r}_i(0) \rvert ^2  \rangle}{6t},
\label{Einstein}
\end{equation}

\noindent where ${\bf r}_i(t)$ represents the center of mass position of ion $i^{\mathrm{th}}$  at time $t$, and the brackets indicate a time average. The self-diffusion coefficient was estimated from the slope of the MSD in the diffusive regime. Unlike simple fluids, the diffusive regime in ILs is reached at fairly long time scales, (see  Figure~\ref{fig:diffusion}), which can exceed correlation times well above 10~ns, for the larger cations (C$_6$MIM or C$_8$MIM). To reach the relevant diffusive regime we integrated the equations of motion using a long time step, 0.01~ps. We ensured that this time step produced stable trajectories and accurate dynamics (\textit{c.f.} MSD obtained with time steps 0.002 and 0.01~ps in Figure~\ref{fig:diffusion}). Using long time step is a definite advantage of CG models, and it has been exploited before in simulations of biological molecules~\cite{doi:10.1021/jp036508g}. All the simulations for the self-diffusion coefficient  were performed at constant volume, with the volume fixed at the average value corresponding to 1~bar and using the v-rescale thermostat with a time constant of 0.2~ps. We generated trajectories spanning 100~ns, in order to ensure the systems reached the diffusive regime. 
The onset of the diffusive regime increases with the cation size, C$_2$MIM$\sim1$~ns, C$_4$MIM$\sim4$~ns, C$_6$MIM$\sim 20$~ns, C$_8$MIM$\sim 60$~ns, at 350~K (see Figure~\ref{fig:diffusion}(a). These time scales are compatible with the relaxation times that can be estimated using standard scaling arguments. For cations with long chain lengths, the self-diffusion coefficient at 350 K  
is rather low, typically $\sim 10^{-11}$~m$^2$~s$^{-1}$, which would give a characteristic time for entering the diffusive regime $\tau >  a^2 / D \approx 25$~ns. This justifies the need to perform very long simulations. 

We report in Figure~\ref{fig:diffusion}(c) the diffusion coefficients obtained for ILs with cations of difference composition at 350 K and 1 bar (additional data at other temperatures are reported in Table 3). In agreement with the experimental observations, the self-diffusion coefficient decreases with the length of the aliphatic chain of the cation; increasing the cation chain length from C$_2$ to C$_8$  the diffusion decreases by a factor of $>20$. The diffusion coefficients obtained with the CG FF compare well with the available  experimental data as well as previous simulation studies using CG models.   It has been noted in experimental studies that  small cations (C$_2$MIM) diffuse faster than the anion, for a variety of anions~\cite{tokuda2005}, while the diffusion coefficients of both ions is essentially the same for larger cations (C$_{>4}$MIM). 
This behavior is  reproduced in our simulations, showing that the diffusion coefficients of both ions  becomes identical within the error of the our computations for  C$_6$MIM (see Figure 5(c).
We expect that the convergence of the diffusion coefficients arises from the structural changes undergone by the IL when the size of the cation increases. Longer chains promote the formation of nanodomains, as discussed in the previous section, with the charges and non-polar regions segregating from each other. For [C$_6$MIM][BF$_4$] the CG predicts negligible differences in the diffusion coefficient. On this basis of the dynamic data we expect that significant segregation in nanodomains takes place between C$_4$MIM and C$_6$MIM. This conclusion is consistent with the evolution of the coarsening of the structure shown in Figure~\ref{snapshots}. 

Additional data for the self-diffusion coefficient predicted by the CG models are collected in  Table~\ref{tab:diff}. The self-diffusion coefficients for small cations at low temperature (298~K) agree well with experimental data, while we find an underestimation for the anions. The results for larger cations also lie below the experimental ones. This behavior is consistent with previous observations, where it was concluded that the level of underestimation increases in highly viscous ILs. 
Indeed, the viscosities increase significantly with cation size (see Table~\ref{tab:diff}), following the experimental observations. The viscosity predictions, considering the difficulties associated to the measurement of this property, are in good agreement with the available experimental data for C$_4$MIM and C$_6$MIM liquids.

Overall, the CG model reproduce well the general dependence of diffusion with temperature and cation size, at similar level of accuracy as those obtained with reparametrized all-atom force-fields\cite{koddermann2007}, which was fitted to reproduce diffusion coefficients. The results represent an improvement with respect previous calculations using a modified version of the force-field developed by Canongia-Lopes et al.~\cite{tsuzuki2009,CanongiaLopes2006}.

\begin{table}[!ht]
 \caption{Self-diffusion coefficients, $D_{_\mathrm{Cation}}$ and $D{_\mathrm{Anion}}$, of C$_\mathrm{n}$MIM-BF$_4$ ILs  and one equimolar mixture, using the CG model discussed in this work, for different temperatures, $T$ in K, and pressure 1~bar. The self-diffusion coefficients are given in $\times 10^{-11}$ m$^2$~s$^{-1}$. The viscosity, $\eta$ is given in mPa~$\cdot$~s. 
The experimental data (Exp.) for C$_2$MIM are taken from  Ref.~\cite{noda} and the remaining data for longer cations from Ref.~\cite{harris}.
\label{tab:diff}} 
 \centering       
 \begin{tabular}{|l|cc|cc|cc|} 
  \hline\hline

  \hline
  
 \small{T}  & \multicolumn{2}{c|}{\small{$D_{_\mathrm{Cation}}$}}  & \multicolumn{2}{|c|}{\small{$D_{_\mathrm{Anion}}$}}  & \multicolumn{2}{|c|}{\small{$\eta$}}   \\ 
    & \small{(CG)} & \small{(Exp.)} &  \small{(CG)} & \small{(Exp.)}  &  \small{(CG)} & \small{(Exp.)}  \\ 
      \hline
      \multicolumn{7}{|c|}{\small{C$_2$MIM BF$_4$}}       \\
  \hline
  \small{298} & \small{5.24$\pm$0.46}              & \small{5.00}                  & \small{2.32$\pm$0.13} & \small{4.20}              &\small{47.20$\pm$2.00} & \small{31.8} \\
   \small{350} &  \small{24.7$\pm$0.30}                &  \small{19.5}       
   &  \small{15.2$\pm$0.50}  & \small{17.0}            &
   \small{10.2$\pm$0.30} &  \small{8.1} \\
    \small{450} &  \small{112.0$\pm$0.30}                &  \small{77.2}                 &  \small{85.41$\pm$1.07}  &  \small{91.6}            &
   \small{2.21$\pm$0.10} &  \small{2.4}\\
  \hline
  \multicolumn{7}{|c|}{\small{C$_4$MIM BF$_4$}}        \\
  \hline
   \small{298} & \small{1.12$\pm$0.11}         &  \small{1.5}                  & \small{ 0.59$\pm$0.10}  &  \small{1.3}   & \small{136.4$\pm$9.1} & \small{109} \\
  \small{350} & \small{12.7$\pm$0.10}                &      \small{9.6}        & \small{9.2$\pm$0.10}  & \small{9.5} & \small{14.8$\pm$0.3 } &\small{15.2}\\
   \small{450} & \small{87.4$\pm$0.30}                & \small{54.47}                & \small{77.51$\pm$0.91}  & \small{67.14}          &
  \small{2.41$\pm$0.12} &   \small{5.8 (388.04~K)~\cite{jacquemin2006density}}\\
  \hline
  \multicolumn{7}{|c|}{\small{C$_6$MIM BF$_4$}}        \\
  \hline
  \small{350} & \small{3.11$\pm$0.05} & \small{6.1} & \small{2.94$\pm$0.05} & \small{6.3} & \small{32.4$\pm$0.90} & \small{23.6} \\
\small{450} & \small{50.61$\pm$0.10} & $-$ & \small{53.30$\pm$0.74} & $-$ & \small{3.41$\pm$0.16} &   \small{13.92 (368.15~K)~\cite{beigi2013investigation}}\\ 
 \hline
  \multicolumn{7}{|c|}{\small{C$_8$MIM BF$_4$}}        \\
  \hline
  \small{350} &  \small{0.72$\pm$0.06} & \small{3.8} & \small{0.89$\pm$0.06} & \small{4.6} & \small{72.9$\pm$10.6} & \small{31.3} \\
  \small{450} & \small{25.67$\pm$0.35} & $-$ & \small{32.54$\pm$0.16} & $-$ & \small{6.72$\pm$1.22}& $-$ \\
  \hline
   \multicolumn{7}{|c|}{\small{[C$_4$MIM]$_{0.5}$ [C$_8$MIM]$_{0.5}$ BF$_4$}}        \\
  \hline
 \small{350} &  \begin{tabular}{c} ${C_4:}$ 3.97 $\pm$ 0.31 \\ ${C_8:}$ 2.22 $\pm$ 0.10 \end{tabular}    
 & $-$ &   \small{2.83$\pm$0.10}       &   $-$      
 & \small{45.9$\pm$8.1} & $-$  \\
  \hline
  \end{tabular}
\end{table}

In light of the convergence of the self-diffusion coefficients of cations and anions between the C$_4$ and C$_6$ chain lengths, the investigation of mixtures including both these two cations, [C$_\mathrm{4}$MIM][BF$_4$]$_{0.5}$--[C$_8$MIM][BF$_4$]$_{0.5}$[BF$_4$] is of particular interest. We show in Figure~5(a) the MSD for the cations and the anions at 350 K and 1 bar, and the diffusion coefficients are collected in Table~3.  The diffusion coefficients of the two cations are different, C$_4$MIM moves faster than C$_8$MIM, and the latter has a self-diffusion coefficient very similar to that of the anion. Our simulations predicts a large modification in the self-diffusion coefficient of the cations and anions for this binary mixture, relative to the corresponding pure fluids,  approximately 3 times increase or decrease for   C$_8$MIM  or  C$_4$MIM.
The self-diffusion coefficient of the anion changes approximately by the same ratio. Following our discussion above regarding the correlation between self-diffusion coefficient and clustering, it stems that the similarity of the diffusion coefficients observed in the mixtures indicates that the liquids features nanodomains  of segregated charged and non-polar regions.
This idea is consistent with the representative structure shown in Figure~\ref{snapshots}(e) where charged and non-polar domains are very well defined.  The characteristic size of the domains as given by the average cluster size (see Figure~4) is similar to that of a  pure liquid with a cation of intermediate size, C$_6$MIM. 


\begin{table}[!ht]
 \caption{Surface tensions of the family of the [C$_\mathrm{n}$MIM][BF$_4$] ILs obtained using the CG model. The experimental data for C$_2$MIM  are taken from Ref.~\cite{souvckova2011surface}, for C$_4$MIM and C$_6$MIM from Ref.~\cite{ghatee2008surface} and for  C$_8$MIM from Ref. ~\cite{restolho2009viscosity}. \label{tab:surftension}} 
 \centering       
 \begin{tabular}{|c|c|c c|} 
  \hline\hline
  Ionic liquid  &  Temperature (K) &  \multicolumn{2}{|c|}{$\gamma$ (mN/m)}   \\ 
   &   & (CG) & (Exp.)   \\ 
  \hline
  \multirow{4}{*}{C$_2$MIM BF$_4$} & 298 &  58$\pm$1   & 53.97\\
  & 350 &  55.4$\pm$0.9 & 50.76 \\
  & 450 & 47.8$\pm$0.6 &  $-$ \\
  \hline
  \multirow{3}{*}{C$_4$MIM BF$_4$} & 298 &  49.8$\pm$3 & 43.6\\
  & 350 &  44.1$\pm$1 & 40.4\\
  & 450 &  38.9$\pm$0.4 & $-$\\
  \hline
  \multirow{3}{*}{C$_6$MIM BF$_4$} & 298 &  {39.3 $\pm$ 4} &39.2\\
  & 350 &  {31.7$\pm$2.2} &36.4\\
  & 450 &  27.5$\pm$0.5 & $-$\\
  \hline
  \multirow{3}{*}{C$_8$MIM BF$_4$} & 298 & $-$ & 32.7 \\
  & 350 & 28.3 $\pm$ 1.4   &30.2 \\  
  & 450 &  19.2$\pm$0.5& 24.5 \\
 
  \hline
  \end{tabular}
\end{table}

\subsection{Interfacial properties}

\begin{figure}[!ht]
  \centering
  \includegraphics[scale=0.5]{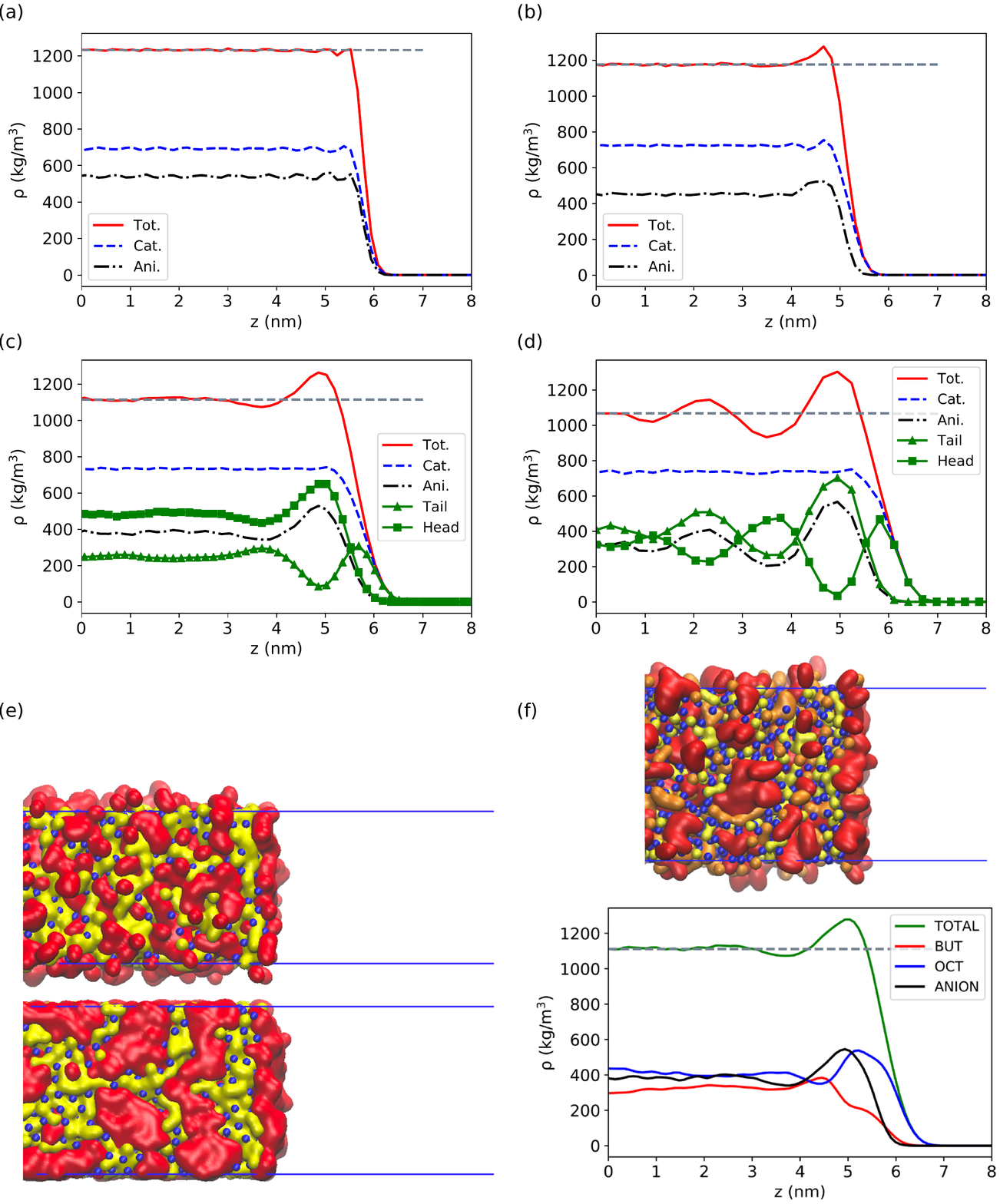}
      \caption{(a) Density profiles of the [C$_{n}$MIM][BF$_4$] ILs, n=2, 4, 6 and 8 ionic liquids at 350~K. The panels (a)-(d) correspond to  n=2--8. The horizontal dashed line in each panel represents the average density of the liquid obtained from NPT simulations at 1 bar pressure. (e) Snapshots of the interfacial region of  C$_{6}$MIM (top) and C$_{8}$MIM liquids (bottom).  The colors, surfaces and spheres have the same meaning as in Figure~4. The blue lines represent the boundary of the simulation box. (f) Density profile (bottom) and snapshot (top) of the mixture [C$_{4}$MIM]$_{0.5}$[BF$_4$]-- [C$_{8}$MIM]$_{0.5}$[BF$_4$]. The colors have the same meaning as in the Figure~4. The surface colored in orange represents the tails of the C$_4$MIM cations. 
      }
    \label{Fig4}
\end{figure}

We investigate now the liquid-vapour interface of the ILs. The surface tension of the interface is determined by the cohesive energy of the liquid, and, therefore, it is a good indicator of the accuracy of the force-field  to captures the correct energy scale of the ionic liquid. Furthermore, it provides information on the ability of the models to reproduce the interfacial behavior, which is important  in electrochemical studies or in extraction and separation in liquid biphasic experiments. 

We computed the surface tension of the liquid-vapor interface through simulations of a liquid slab, which was generated by replicating a pre-equilibrated configuration obtained from a previous NPT simulation,  3 or 4 times along the $z$ direction, normal to the interface. The liquid slab was then surrounded by vacuum regions and equilibrated for several nanoseconds before computing averages.  Additional details are provided in the Model and Methods section.
The surface tension, $\gamma$, was calculated by using the pressure tensor through the following equation:

\begin{equation}
\gamma = \frac{L_z}{2}\biggr[  \langle P_{zz}\rangle  -  \frac{ \langle P_{xx}\rangle - \langle P_{yy}\rangle }{2} \biggl],
\label{tension}
\end{equation}

\noindent where $P_{\alpha \beta}$ are the pressure tensor components in the direction $\alpha$ and $\beta$ ($\alpha$ and $\beta$ = $x$, $y$ or $y$), and the brackets, $<>$, indicate a time average. $<P_{zz}>$ is the pressure component perpendicular to the interface plane, and its value is determines by the vapor pressure of the liquid. Our simulations showed a zero density in the vapor region, even at the highest temperatures simulated here. This result  is consistent with the very low vapor pressures observed in ILs. 

\begin{figure}[!ht]
  \centering
    \includegraphics[scale=0.75]{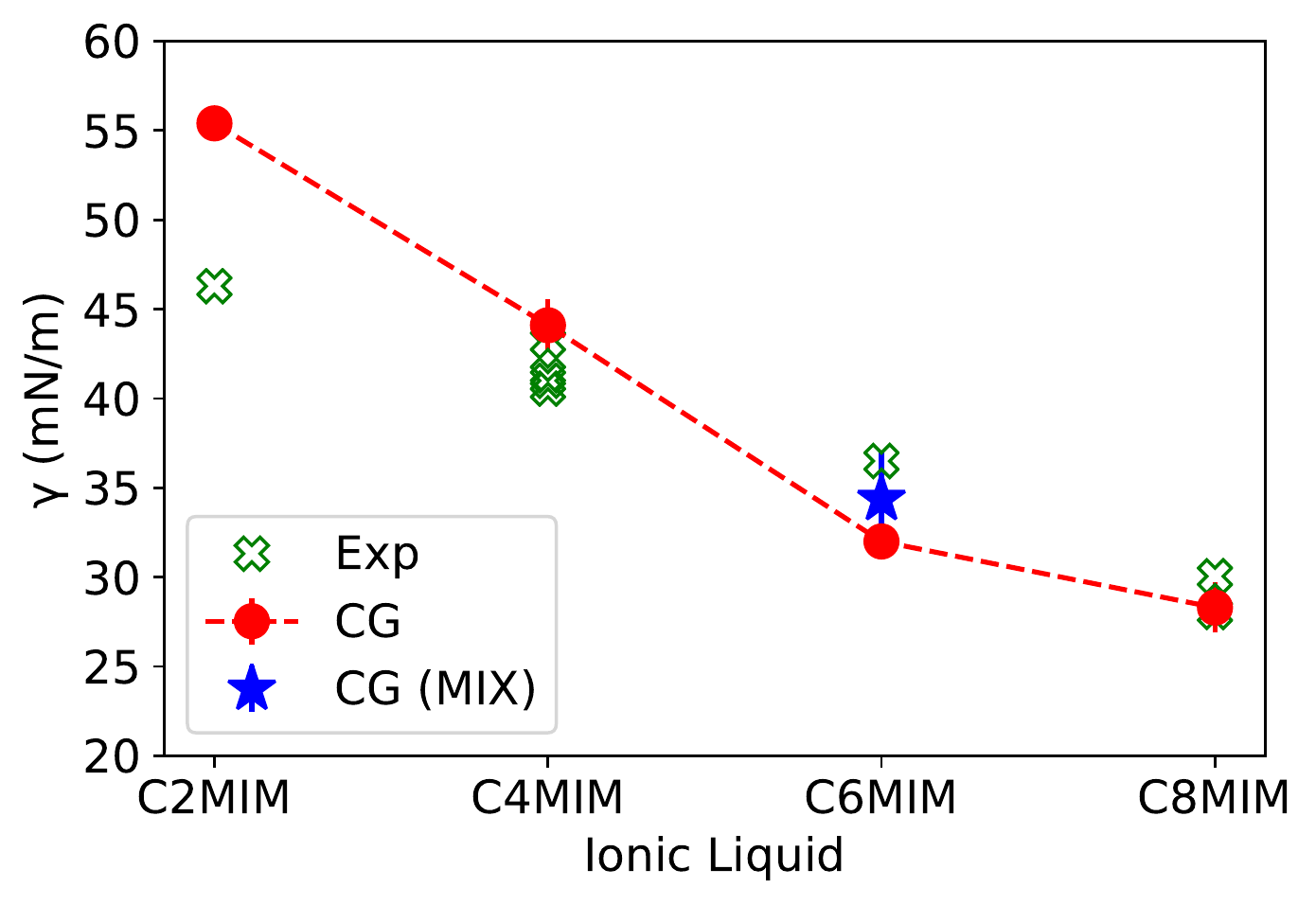} 	
      \caption{Surface tension of the ILs as a function of the cation size at 350~K. The crosses represent experimental data from Ref. \cite{SHAMSIPUR201043,ghatee2008surface,doi:10.1021/je800710p,RESTOLHO200982}. The circles represent simulation data obtained in this work using the CG model. The star represents the simulation data for the mixture [C$_{4}$MIM]$_{0.5}$[BF$_4$]-- [C$_{8}$MIM]$_{0.5}$[BF$_4$] . The dashed line is a guide to the eye.}
    \label{Fig7}
\end{figure}

We show in Figure~6 representative density profiles for the [C$_\mathrm{n}$MIM][BF$_4$] family of ILs at 350 K. The formation of nanodomains observed in bulk is also evident for interfacial systems. There is clear preference for  cations  with long aliphatic chains to populate the interface, resembling the behavior observed in surfactant molecules. The adsorption induces strong layering in the direction perpendicular to the interface plane, characterized by  alternating charged/non-polar regions that penetrate deeply into the bulk, $>5$~nm for the larger cations, C$_8$MIM (see oscillations in antiphase of the tail and head contributions). The analysis of the oscillatory profiles renders a characteristic wavelength of the order of 3~nm, which we interpret as measure of the typical size of the domains formed in [C$_8$MIM][BF$_4$]. The layering is also evident in C$_6$MIM IL, but the oscillations induced by the liquid-vapor interface decay much faster. We observe a weak maximum for the smaller cations in the series, C$_2$MIM and C$_4$MIM, which is consistent with the lack of significant nanodomains in these liquids (see Figure~4). 

 The surface tension of the larger cations should be determined to a larger extent by the surface energy of the aliphatic chains, and, consequently, the surface tension should decrease when considering IL with smaller cations, where the ionic interactions play a more important role. This notion is clearly demonstrated in Figure~7. Indeed, the CG predicts 
 a reduction in the surface tension with increasing cation size. The surface tension results are in good agreement with  the experimental data. The results for [C$_2$MIM][BF$_4$] feature a larger deviation with respect to experiments. Interpolation of our data  at 400~K  gives a surface tension of $\sim$ 51~mN~m$^{-1}$, which is higher than the value reported before for the rigid version of this CG model, 44.3~mN~m$^{-1}$~\cite{Merlet2012}. There are two differences between that model and the one employed in our work. Firstly, the cut-off, 1.5~nm \cite{private} {\it vs} the full potential (no cut-off) employed here, and secondly we model the cation as a flexible molecule with rigid bonds, while the model in ref.~\cite{Merlet2012} is fully rigid. Repeating our simulation with a cut-off of 1.5~nm at 400 K we get  $44.1\pm1.4$~mN~m$^{-1}$, in very good agreement with the 
 results reported before. Hence, we conclude the origin of the difference is the dispersion interactions, and the flexibility associated to angular terms is not introducing significant dissimilarities.
 
 
The surface tension of the mixture [C$_4$MIM]$_{0.5}$[BF$_4$]--[C$_8$MIM]$_{0.5}$[BF$_4$] lies in between those of the corresponding pure liquids, and it is very close to the surface tension of  [C$_6$MIM][BF$_4$], hence following again behaviors similar to that reported 
for structural and dynamic properties. The interface is enriched in the C$_8$MIM cation, which protrudes more towards the vapor phase. The analysis of the densities and simulations snapshots (see Figure~6) confirms this idea. 

\section{Conclusions}
We have investigated the properties of the family of [C$_\mathrm{n}$MIM][BF$_4$] ILs $\mathrm{n}=2...8$, building a new coarse grained force-field based on an earlier model by Merlet et al. \cite{Merlet2012}. To model the aliphatic chains of the cations we used a 2 to 1 mapping strategy, where 6 (-CH$_2$-CH$_2$-) or 7  (-CH$_2$-CH$_3$) atoms are mapped into one pseudo-atom. 
This mapping into a smaller number of atoms and interactions, speeds up the simulation  significantly, allowing the use of long time steps $0.01$~ps, with the possibility of performing $\mu$s time scale simulations very efficiently. These long simulations are essential to describe the slow dynamics of the ILs involving larger cations. Our primary motivation in exploring these force-fields is their use in future investigations of confinement effects, such as in lubrication processes, hence the enhanced time scales should prove helpful.

The CG models reproduce well the equation of state of the liquid at 1 bar, as well as the main structural properties of ILs. The radial distribution functions agree  with those obtained with well established full atomistic models, OPLS-AA~\cite{CanongiaLopes2006}. They reproduce accurately the characteristic oscillatory behavior of the radial distribution function, which is dominated, even for the larger cations, by electrostatic correlations. 

ILs consisting of large cations structure themselves in nanoscale regions with segregation into charged and non polar domains. This observation highlights the surfactant like nature of ILs, and it is consistent with previous studies, as well as proposals on the formation of micelle like and bilayer structures\cite{kornyshev2007}.   Our cluster analysis indicates that significant nanoscale structuring starts to take place between C$_4$MIM and  C$_6$MIM. The change in the  IL structure is reflected in the self-diffusion coefficient of the ILs. Nanostructuring drives the convergence of the diffusion coefficients of cations and anions as the size of the cation increases. For system with significant nanostructuring, C$_{\ge 6}$MIM the diffusion coefficients of both ions is identical within the uncertainty of our computations. These observations are in good agreement with  experimental studies using NMR probes, showing that, despite their simplified energy surfaces, CG models can be successfully used to study dynamic properties. Indeed, the predictive power of these models is similar, or in some cases better, than that of fully atomistic force-fields. 

We  tested the transferability of the CG model by studying IL mixtures. Simulations of equimolar mixtures of ILs, [C$_\mathrm{n}$MIM][BF$_4$] ILs $\mathrm{n}=4 \mathrm{~and~} 8$, indicate that the mixing is nearly ideal. This result is also consistent with experimental studies of excess volumes for this IL.

The dependence of the surface tension of the liquid-vapor interface with temperature is consistent with experimental data. The break in symmetry induced by the liquid-vapor interface induces lateral structure in the IL. Charged and nonpolar domains are observed in the direction normal to the liquid-vapor interface. These domains are reminiscent of the nanostructures formed in bulk. The lateral structure  is driven by the enrichment of the non-polar regions of the cations at the liquid surface, which minimize the surface free energy.  This observation also highlights the surfactant like characteristics of large imidazolium cations.

Overall, the CG model presented here describes relevant  structural, dynamic and interfacial properties of [C$_\mathrm{n}$MIM][BF$_4$] ILs. These properties  are in good agreement with available experiments or full atom simulations. Thus, the CG model provides a good starting point for further developments to include other ILs, as well as alkyl chains with an odd number of carbon atoms, which were not considered here. However, the set of ILs considered here already features significant structural changes driven by the increasing importance of the non-polar regions of the molecule. The impact of these structural features in electrotuneable lubrication is of great interest, and its study will be reported in a forthcoming article.

\section{Acknowledgements}
We would like to acknowledge the award of a Research Grant by The Leverhulme Trust (RPG-2016-223). We thank Michael Urbakh and Alexei Kornyshevs for many illuminating discussions, and  
M. Salanne for providing information on the cut-off employed in Ref. 15.

\bibliography{ref}
\end{document}